\documentclass[a4paper,12pt]{article}
\usepackage{amsmath}
\usepackage[psamsfonts]{amssymb}
\usepackage[mathscr]{eucal}
\begin{document}

\begin{center}

\section*{The system of the  vortex-like structures: \\ the viewpoint on a turbulence modelling }

\vspace{3mm}

{ S.V. Talalov }

\vspace{3mm}

  {\small Dept. of Applied Mathematics,   Togliatti State University, \\ 
 14 Belorusskaya str.,  Tolyatti,     445020 Russia.     }\\
svt\_19@mail.ru

\end{center}

\begin{abstract}
In this study   we suggest new approach to turbulence modeling.  To develop this approach,  we construct the set of the vortex-like dynamical systems evolving in the  space $E_3$. These systems  are constructed using the AKNS hierarchy so that they   generalize  thin closed vortex filament  described in the local induction approximation.
 We construct and investigate the Hamiltonian dynamics of these   systems  in terms of the non - standard variables. 		We formulate the principles of quantization these systems focusing on the quantum description of the filaments  with linearized dynamics.  To describe the energy of vortex-like structures, we apply a new method based on the use of the central extended Galilei group. In a special case, we calculate the energy levels of a system of quantized vortices.
		The proposed approach makes it possible to describe  	both	 the interaction of such vortex - like filaments 		and the processes of their  creation  and annihilation in a natural way. 
		As a result, we have constructed a quantum turbulent flow model that demonstrates the relationship between stochastic and integrable dynamics.		
\end{abstract}

{\bf keywords:}
vortex filament quantization;  AKNS hierarchy;  extended Galilei group;  turbulence modelling.  
 

\vspace{5mm}

	 \paragraph~
	
	\section{Introduction}
	
	~~~The turbulence, both classical and quantum, continues to be a phenomenon that is not fully understood. A common approach to describing this complex phenomenon is to consider a turbulent flow as a tangle of vortex structures. 
A large number of works are devoted to this issue. 
{ For the first time, the description of a turbulent medium as a tangle of vortices was proposed in Feynman's work \cite{Feyn}.
The achievements in this matter, which followed over several decades, were summarized, for example, in the books \cite{Donn,Frisch}. 
Despite many years of studying turbulence, this complex phenomenon is still of interest to researchers. Indeed, ''\dots the understanding of turbulent flows is one of the biggest current challenges in physics'' \cite{PoMuKr_1}.
Quantum effects add complexity to the study of this problem. So, the differences between the classical and quantum turbulence were considered in the paper  
\cite{Vinen}. Along with the theoretical understanding of the problem, numerical studies were also carried out \cite{Aarts} (see also \cite{TsFuYu,Fuka} for resent results here).

As usual, the theoretical study of the phenomenon involves building some kind of model.
The Gross-Pitaevskii model is the most used theory in modeling quantum turbulence.
So, one of the predictions of this theory is the values of quantum circulation $\Gamma_n \propto \hbar n$, where the numbers $n$ are integers. The study of the statistics of the distribution of the $\Gamma_n$ value  in a turbulent flow was investigated in work \cite{MuPoKr} by numerical simulation.
The Hall-Vinen-Bekharevich-Khalatnikov (HVBK) model is also  widely used for numerically study quantum turbulence. For example, this approach was used for this purpose in the paper \cite{ZDLD}.

Currently, it is considered an established fact that turbulent flow is characterized by a wide range of length and time scales.
This question was discussed, for example, in the paper
 \cite{Gourianov}.
Topological and geometrical aspects of a quantum turbulent flow was studied in the work \cite{Barren}.  More complicated problems were also considered: quantum turbulence in a channel with an  counterflow of superfluid turbulent helium was studied in the paper \cite{Nemir}.

Despite the large number of works in the field of quantum turbulence, this phenomenon cannot be considered fully investigated. In this study, the author develops  new approach to the turbulent flow modeling.    This approach  demonstrates the relationship between the regular and stochastic dynamics of vortex-like filaments in a tangle. Moreover, it makes it possible in principle to calculate a partition function for a turbulent flow in certain cases. }

\section{The hierarchy of the vortex-like dynamical systems}

  Nevertheless, the following question would be appropriate  in a framework  of  any   approach: how to describe a single vortex in  a tangle? Corresponding description will be necessary, for example, when constructing a quantum theory of such  complex structures.
Approaches that have been developed to describe of a single thin vortex filament $ {\boldsymbol{r}}(t ,s)$  are well known. One of the most common ways of such a description is to use the Local Induction Approximation. Corresponding equation (LIE) is written as follows::
\begin{equation}
\label{LIE}
        \frac{ \partial {\boldsymbol{r}}({t^\ast} , s)}{\partial {t^\ast}} ~=~ A\,
       \frac{ \partial{\boldsymbol{r}}({t^\ast} ,s)}{\partial s}\times \frac{\partial^{\,2}{\boldsymbol{r}}({t^\ast},s)}{\partial s^2}\,, 
\end{equation}
We use    notation  $s$  for    the  natural   curve  parameter      and notation
 $t^\ast = {t\Gamma  }/{4\pi}$   for the ''time'' parameter. Symbols $\Gamma$ and $t$ mean the circulation and real time correspondingly. 
After certain assumptions \cite{AlKuOk}, a vortex filament    with a nonzero flow inside the core\footnote{We assume that the  core radius is small but non-zero.} is described by the equation:

\begin{eqnarray}
        \label{LIE_dim}
        \frac{ \partial {\boldsymbol{r}}({t^\ast} , s)}{\partial {t^\ast}} &=& A\,
       \frac{ \partial{\boldsymbol{r}}({t^\ast} ,s)}{\partial s}\times \frac{\partial^{\,2}{\boldsymbol{r}}({t^\ast},s)}{\partial s^2} + \nonumber\\[2mm]
				~~ & + &  B\left(\frac{\partial^{\,3}{\boldsymbol{r}}({t^\ast},s)}{\partial s^3} + 
        \frac{3}{2}\,\biggl\vert\, \frac{\partial^{\,2}{\boldsymbol{r}}({t^\ast},s)}{\partial s^2}\biggr\vert^{\,2}\frac{ \partial{\boldsymbol{r}}({t^\ast} ,s)}{\partial s}\right)\,.
				        \end{eqnarray}

Of course, the application of equations (\ref{LIE}) and (\ref{LIE_dim}) to describe individual vortex filaments in a tangle is impossible.  Indeed, the influence of other vortices in the vortex-tangle cannot be ignored here.
On the other hand, it is also impossible to take into account the real physical influence  a lot of nearby random vortices on a particular filament in a turbulent flow.  Most probably,  this problem is  no less difficult than solving the Navier-Stokes equations in a general case.
In this regard, the following question arises: can we modify equations (\ref{LIE}) and (\ref{LIE_dim}) so that the dynamics of the filament is described by a sufficiently large number of parameters?
In this case, the influence of neighboring vortices can be taken into account indirectly, through one or another distribution of these parameters. With this approach, we save the opportunity to describe the dynamics (classical and quantum) of a separate vortex filament.

The purpose of our study is as follows. We  develop a theory that can be considered as a mathematical model of a tangle of vortex-like filaments. This model will include the following items:
\begin{enumerate}
\item  We construct the dynamical theory of the separate vortex-like structure in Hamiltonian form.
This theory  generalizes  dynamics (\ref{LIE}) and (\ref{LIE_dim}) in some way. 
To solve this problem, we define the Hamiltonian structure of such systems in terms
of non-standard  fundamental  variables. So, these variables are divided into two groups: internal and external ones; 
\item To quantize these systems, we linearize the dynamics of considered objects. We will also discuss the relevance of this procedure in our model;
\item We construct the quantum theory of introduced  vortex-like filaments.
 The constructed model makes it possible  to describe the interaction of the vortex-like structures at  quantum level by means of    the   many-body  theory  methods;
\item In a sense, the theory combines elements of integrability and stochasticity in the behavior of the objects under consideration.
\end{enumerate}

For our further studies,, we use a representation for  smooth closed filament of length $S$   in the form
\begin{equation}
        \label{proj_r}
                                   {\boldsymbol r}(\tau,\xi) ~=~  {\boldsymbol{q}} + 
         {R} \int\limits_{0}^{2\pi}  \left[\, {\xi - \eta}\,\right] {\boldsymbol j}(\tau,\eta) d\eta\,, \qquad  R = \frac{S}{2\pi} \,, \qquad \xi = \frac{s}{R}\,,
                  \end{equation}
where dimensionless evolution parameter $\tau = {t\Gamma  }/{4\pi R^2}$ was introduced. The 
vector ${\boldsymbol{q}} = {\boldsymbol{q}}(\tau)$ defines the position of the filament in the coordinate system (for example,  the   center of the circular vortex ring).
The notation $[\,x\,]$ means the integer part of the number $x/{2\pi}$:
\begin{equation}
        \label{int_part}
 [\,0\,] = 0\,, \qquad  [\,x+2\pi\,] = [\,x\,]+1\,, \qquad \forall\,x\,.
\end{equation}
 Vector ${\boldsymbol j}(\tau,\eta)$ is the unit  tangent (affine) vector for this filament. 
Let the vector  ${\boldsymbol r}(\tau,\xi)$ satisfies   equation (\ref{LIE}).
Then, the function ${\boldsymbol j}(\tau,\eta)$ satisfies the equation for continuous Heisenberg spin chain:
\begin{equation}
        \label{CHSCeq}
        \partial_\tau {\boldsymbol{j}}(\tau ,\xi) ~=~ 
   \alpha {\boldsymbol{j}}(\tau ,\xi)\times\partial_\xi^{\,2}{\boldsymbol{j}}(\tau,\xi)\,.
        \end{equation}
	It is well known that this equation is gauge equivalent to the 		non-linear Schr\"odinger equation 		 \cite{TakFad}.   The connection between the LIE  and the nonlinear Schrodinger equation was firstly established in the  work 	\cite{Hasim}.		
This equivalence makes it possible to construct new dynamical systems in space $E_3$ that generalize dynamical systems (\ref{LIE}) and (\ref{LIE_dim}).
 Despite the formal definition, such systems, according to the author, open up a new point of view on a simulation    of vortex filaments  behavior in a vortex tangle.

First, we will consider the Hamiltonian description of systems (\ref{LIE}) and (\ref{LIE_dim}) in terms of non-standard variables. Such a description was proposed in the authors' work
\cite{Tal_18}.  The main points of the suggested approach   are following.
\begin{itemize}
\item We include the circulation $\Gamma$ as an additional independent variable in our theory.
Because this postulate,  we take into account the movement of the surrounding   fluid.  Indeed,  the equations (\ref{LIE}) and (\ref{LIE_dim})    describe some formal dynamics of the curve  ${\boldsymbol r}(\tau,\xi)$ only. 
Therefore, we describe the dynamics of a filament  by 
the following set  ${\cal A}$ of the independent variables: 
$$  {\cal A} ~=~ \{\, \Gamma\,, \boldsymbol{q}\,, \boldsymbol{j}(\xi)\, 
 \}\,,$$
\item We replace the set ${\cal A}$ by the set ${\cal A}^{\prime}$, where
\begin{equation}
\label{new_set}
{\cal A}^{\,\prime} ~=~ \bigl\{\, {\boldsymbol p}\,,  {\boldsymbol{q}}\,;  {\boldsymbol j}(\xi) \,\bigr\}\,.
 \end{equation}
The momentum ${\boldsymbol p}$ is connected with variables $\Gamma$ and $\boldsymbol{j}(\xi)$ by means of standard hydrodynamical formula 
\begin{equation}
        \label{p_and_m_st}
        {\boldsymbol p} ~=~ \frac{\varrho_0}{2 }\,\int\,\boldsymbol{r}\times\boldsymbol{w}(\boldsymbol{r})dV\,,
        \end{equation}
where  $\varrho_0$ is a fluid density and vector-function    $\boldsymbol{w}(\boldsymbol{r})$  means  the vorticity of the vortex filament.   This function  is calculated as
\begin{equation}
        \label{vort_w}
     {\boldsymbol{w}}(\boldsymbol{r}) ~=~  \Gamma
                  \int\limits_{0}^{2\pi}\,\hat\delta(\boldsymbol{r} - \boldsymbol{r}(\xi))\partial_\xi{\boldsymbol{r}}(\xi)d\xi\,.
       \end{equation}
 
\item  The appearance of the momentum components  $p_i$  as independent variables 
						 makes it possible  to define the Galilei boosts  in a natural way.
						As a consequence, we assume  that 	the central extended Galilei
						group\footnote{We use one parameter  central extension of this group.  The mass constant $m_0$ (central charge) is the parameter of this extension. } $\tilde{\mathcal G}_3$  is   the space-time symmetry group in our model.  Therefore, we apply 				
the group-theoretical approach to the definition   of the energy of thin  vortex filament.
	\item 			The structure of  Hamiltonian is motivated by the expressions for Cazimir functions of Lie algebra of the  group  $\tilde{\mathcal G}_3$:			
	\begin{equation}
        \label{H_general}
	{H}_0(p_1,p_2,p_3\,;{\boldsymbol j}) ~=~ \frac{1}{2m_0}\sum_{i=1}^3 p_i^{\,2}   +  
	 H_{int}[\,{\boldsymbol j}; \alpha_1,\dots,\alpha_n\,]\,. 
	\end{equation}						
	The term 	 $H_{int}[\,{\boldsymbol j}; \alpha_1,\dots,\alpha_n\,]	$		provides dynamics of the variables 		${\boldsymbol j}(\xi)$ that depends on some parameters 
	$\alpha_1,\dots,\alpha_n$.						 

\end{itemize}

The formulas (\ref{p_and_m_st}) and (\ref{vort_w}) are written in a compact form:
\begin{equation}
        \label{impuls_def}
    {\boldsymbol{p}}  ~=~    \varrho_0 {R}^2 \Gamma        {\boldsymbol f} \,, 
							\qquad 
						{\boldsymbol f} ~=~ \frac{1}{2}\iint\limits_{0}^{2\pi}  \left[\, {\xi - \eta}\,\right]\,{\boldsymbol j}(\eta)\times{\boldsymbol j}(\xi)d\xi  d\eta\,.	
      \end{equation} 
Please note that the replacement $\Gamma \to {\boldsymbol p}$ leads to the certain constraints on the set ${\cal A}^{\,\prime}$ in general. 
We will not discuss this issue here \cite{Tal_EJM}.

The theory has three natural  dimensional constants  that are relevant to the physical system being described. These constants are: the fluid's density $\varrho_0$, the speed of sound in this fluid $v_0$ and the mass constant $m_0$. 
In order to simplify some formulas,  we will use the auxiliary  constants\footnote{Note that  there may be other natural dimensional constants in theory: for example, the radius $R_1$ of the pipe in which the vortex tangle in question moves.}
$R_0 = \sqrt[3\,]{m_0/\varrho_0}$,
 $~t_0 = R_0/v_0$ and   ${\cal E}_0  = m_0 v_0^2$ along with constants $\varrho_0$, $v_0$ and $m_0$.

The Poisson bracket structure was postulated as following:
	 	\begin{eqnarray}
  \{p_i\,,q_j\} & = & \,\delta_{ij}\,,\qquad i,j =1,2,3\,, \nonumber \\
  \label{ja_jb}
  \{ j_a(\xi), j_b(\eta)\} & = & - (2/R_0 m_0 v_0)\, \epsilon_{abc} j_c(\xi) \hat\delta(\xi- \eta)\,, \qquad      \epsilon_{123} = 1\,.
  \end{eqnarray}  
	All other brackets  vanish. 
	In accordance with the definition (\ref{ja_jb}), the function  ${\boldsymbol{j}}^{\,2}(\eta)$ annulates the brackets of the fundamental variables. Thus, the 
	constant ${\boldsymbol{j}}^{\,2}(\eta)$ numbers the symplectic sheets in the phase space of our dynamical system. 	
	  Next, we consider a single symplectic sheet with ${\boldsymbol{j}}^{\,2} =1$.

  To develop our approach  it let us note that the equation  (\ref{CHSCeq}) lead to   the infinitely series of the 
	integrals of the motion $I_n = I_n[\,{\boldsymbol j}\,]$, $n=0,1,2,\dots$ \cite{TakFad}.  These integrals generate  Hamiltonian flows that lead to the hierarhy of the non-linear integrable differential equations,
	 which is known   as hierarhy AKNS \cite{AKNS}.   
	So, we use the  Hamiltonian 
	\begin{equation}
        \label{H_LIE}
	 H_{int}  = {\mathcal E}_0  \int_{0}^{2\pi} ( \partial_\xi {\boldsymbol j}(\xi))^2 d\xi = 4 {\mathcal E}_0   I_1  
	\end{equation}
to generate the dynamics in accordance with equation (\ref{CHSCeq}). As for the full Hamiltonian  (\ref{H_general}), it provides the dynamics of the filament according to the equation (\ref{LIE}).

Let's now add to expression (\ref{H_LIE}) the term that is proportional to the integral of motion $I_2$. Then the resulting full  Hamiltonian 
\begin{eqnarray}
        \label{H_I1_I2}
~~&~&~~	{H}_0(p_1,p_2,p_3\,;{\boldsymbol j}) ~=~ \frac{1}{2m_0}\sum_{i=1}^3 p_i^{\,2}   ~+\nonumber \\[3mm] 
	~&+&  {{\mathcal E}_0}\,\left(\alpha_1\int\limits_{0}^{2\pi} \bigl( \partial_\xi {\boldsymbol j}(\xi)\bigr)^2 d\xi
+ \alpha_2 \int\limits_{0}^{2\pi}  \partial_\xi {\boldsymbol j}(\xi)\bigl({\boldsymbol j}(\xi)\,\times\,\partial^{\,2}_\xi {\boldsymbol j}(\xi) \bigr) d\xi\right)
	\end{eqnarray}	
will provide the dynamics according to the equation (\ref{LIE_dim}).

  Considered Hamiltonians $H_{int}$    have a natural geometrical interpretation. Indeed, taking into account  Frenet-Serret equations for the curve (\ref{proj_r}),	we can find that following simple formula holds:
								
	$$ H_{int}[{\boldsymbol j}; \alpha_1,\alpha_2] = 
 {{\mathcal E}_0}\,\left(\alpha_1 \int\limits_{0}^{2\pi}  \bigl(k(\xi)\bigr)^2 d\xi
+ \alpha_2 \int\limits_{0}^{2\pi}  \bigl(k(\xi)  \bigr)^2 \varkappa(\xi) d\xi\right) \,. $$
	The function $k(\xi)$ is the curvature of the curve ${\boldsymbol{r}}(\cdot,\xi)$  here and the function $\varkappa(\xi)$ is the torsion of this curve.

The structure of the Hamiltonian (\ref{H_I1_I2}) makes it natural to use AKNS hierarchy    to generalize the  dynamic system (\ref{LIE_dim}).
Let the number  ${\sf N}>0$ is the natural number.
To describe the hierarchy of the vortex-like filaments, we consider the
 Hamiltonian (\ref{H_general}), where
	
	\[ H_{int}[\,{\boldsymbol j}; \alpha_1,\alpha_2\,] ~\to~ H^\varkappa_{int}[\,{\boldsymbol j};\, \alpha_1, \dots,\alpha_{\sf N}\,]\,, \]
	where
	\begin{equation}
	\label{New_H}
			H^\varkappa_{int}[\,{\boldsymbol j}; \alpha_1, \dots,\alpha_{\sf N}\,]  ~=~ 4{\mathcal E}_0\sum_{k=1}^{\sf N} \alpha_k I_{k}[\,{\boldsymbol j}\,]\,, \qquad \alpha_j \in {\sf R}  \,. 
			\end{equation}
	
	The certain number of  constants $\alpha_k$  may be zero. The fact that $\alpha_{k_1}=0$ for some index $k_1$ means that the corresponding integral of motion  $I_{k_1}[\,{\boldsymbol j}\,]$  does not affect the dynamics of the filament.  To account for this effect,  we define the real number
$ \varkappa \in [0,1]$ that is written in binary form as
\begin{equation}
\label{number_kappa}
 \varkappa ~=~ 0,b_{{\sf N}-1}b_{N-2}\dotsb_{1}b_{0}\,0\,0\,0\,0\,0\,0\,0\dots\,. 
\end{equation}
The numbers $b_{k-1}$ are mapped to  constants $\alpha_k$ according to the following rule:  $b_{k-1} = 0$ for the case $\alpha_k =0$  and $b_{k-1} = 1$   otherwise . 
 	The number     $\varkappa$ can take $2^{\sf N}$ values. This number determines the type of vortex-like filament evolution  of concrete vortex loop in a vortex tangle.

		Therefore, we have the hierarchy of the dynamical  systems  in framework of developed theory. These systems generalize the vortex systems (\ref{LIE}) and (\ref{LIE_dim}) naturally. In general, the dynamics of the vortex-like filament (\ref{proj_r}) is described by ${\sf N}+1$-order non-linear differential equation:
		\begin{equation}
\label{dyn_eq_g}
		 \partial_\tau{\boldsymbol r} ~=~ \{H^\varkappa_{int}[\,{\boldsymbol j}; \alpha_1, \dots,\alpha_{\sf N}\,]\,,{\boldsymbol r}\}\,. 
		\end{equation}
		
		In conclusion of this section, we will say a few words about the geometric interpretation of such dynamical systems. 		
	First, 	the functions $P_n(\xi)$ in the integrals $I_n =\int_0^{2\pi}P_n(\xi)d\xi$ are polynomials of the solution $\psi(\xi)$ of the 	non-linear Schr\"odinger equation and corresponding derivatives  $\partial^n\psi(\xi)$. 
		 Second, the function $\psi(\xi)$  can be reconstructed on the curvature
	$k(\xi)$ and the torsion  $\varkappa(\xi)$ in accordance with Hasimoto results \cite{Hasim}. 		Therefore, the Hamiltonians $H^\varkappa_{int}[\,{\boldsymbol j}; \alpha_1, \dots,\alpha_{\sf N}\,]$  will be functions of quantities	$k(\xi)$ and $\varkappa(\xi)$.

	\section{Quantization}

			~~~The problem of quantum description of vortices has a very long history.
			The author will not take the responsibility to give a full review of the literature on this issue here. In most of the works that have been cited above, this problem was also discussed in some way.					
			  In addition, it is  necessary to mention the book \cite{Sonin}, which also has a detailed review of the literature on the problem.
		The quantum Hasimoto transformation LIE $\to$ non-linear Schr\"odinger equation  was studied in the work \cite{Gorder} recently.

		We emphasize that the general dynamics of the  vortex filaments is described by nonlinear equations, even for simplified case of the  local induction approximation.	 
				The quantization of the corresponding non-linear integrable  models is complex problem (see, for example,  	\cite{TakFad_Q,Sklyanin}). 
	According to our initial assumptions, the discussed  vortex-like  dynamical systems  form some ''entangled  structure''  that models the turbulent flow. 
	The lifetime of vortices in a turbulent flow is small; this value  restricted by the period
			$T = 2\pi/W$, where symbol $W$ means the vortisity  \cite{Tenn}.
		Thus, it seems appropriate to consider the quantization of small perturbations of vortex loops.	Such perturbations are described by the linear equations at corresponding approximation. Therefore, the quantization of these perturbations  can be performed using standard methods. The main question here is follows:   what   fundamental variables must  parametrize the phase space of the dynamical systems under consideration?
		It is well known that quantization of  variables that are  canonically equivalent in a classical theory,  leads to different results at the quantum level\footnote{
For example, the  quantization of  such a simple canonical system as  harmonic oscillator in terms of action-angle variables demonstrates many unexpected and interesting results \cite{Kast}.}.			
	Despite more than a century of development of quantum theory, the following  
 Dirac's words are still relevant: 					{\it''\dots methods of quantization are all of the nature of practical rules, whose application depends on consideration of simplicity''} \cite{Dirac}.
	
	The quantization of small oscillations of the vortex ring was suggested in the authors' article   \cite{Tal_EJM} in terms of non-standard Hamiltonian variables  (\ref{new_set}).
	As special  result, the set of acceptable values of quantized circulation $\Gamma_{[n]}$  was calculated    within the framework of this approach \cite{Tal_PoF22}.  As it turns out, this set  is wider than the standard one $\Gamma_n \propto n$, where $n$ is natural number.
		The developed theory allows us to interpret a quantized vortex as a certain particle with an internal structure. 	Consequently, we can  apply the methods of the theory of many bodies to describe the processes of the interaction, creation and annihilation  of the vortices \cite{Tal_PoF23}.

	Let us make  the overview of our quantization scheme.		
		First, we consider the single closed  vortex-like filament 	 that has some fixed shape.
	This filamet is  homotopically equivalent to a circle and is  characterized by a certain tangent vector ${\boldsymbol j}_0(\eta) = {\boldsymbol j}_0(\tau,\eta)$:
	$$	{\boldsymbol j}_0(\eta + 2\pi)	= {\boldsymbol j}_0(\eta)\,, \qquad  
			{\boldsymbol j}_0^2(\eta) = 1\,.$$
			We consider the small perturbation   of this tangent vector: 
 		\begin{equation}
        \label{tang_v_per}          																
		\boldsymbol{j}_0(\tau,\xi)  ~\to~ {\boldsymbol{j}_0}(\xi) + \varepsilon {\boldsymbol{j}}_p(\tau,\xi) \,,	\qquad 		\varepsilon << 1\,.													
\end{equation}
	Additionally, we  suppose that these  excitations be transverse:
\begin{equation}
        \label{iden_1}
				{\boldsymbol j}_p(\tau ,\xi){\boldsymbol{j}_0}(\xi) ~\equiv~ 0\,
	\end{equation}
The small perturbation of the circular vortex filament  without the restriction (\ref{iden_1}) has been studied in the work \cite{Ricc}.
	Various aspects of the theory related to vortex ring oscillations have also been studied in the works \cite{Kop_Chern,Kik_Mam}.  Besides that,  the small perturbations of straight vortex filaments were studied in the work \cite{Majda}.

	The condition (\ref{iden_1}) leads to decomposition
\begin{equation}
        \label{j-two}
{\boldsymbol j}_p(\tau ,\xi)   
  ~=~   j_\rho(\tau,\xi){\boldsymbol{e}}_\rho  +       j_z(\tau ,\xi)  {\boldsymbol{e}_z}\,,
\end{equation}
where the vectors $\{ {\boldsymbol{e}_\rho}\,,    {\boldsymbol{j}_0}\,, {\boldsymbol{e}_z}\,\}$ denote the local orthonormal basis in the corresponding point of the loop.
		For convenience, we  define the  complex-valued function
	\[  {\mathfrak J}(\tau ,\xi)  ~=~  j_\rho (\tau ,\xi) + {\rm i}  j_z(\tau ,\xi)\,.\]
	This function allows us define the complex amplitudes ${\mathfrak j}_{\,n}(\tau)$:
		\begin{equation}
        \label{sol_general}
				{\mathfrak J}(\tau ,\xi) ~=~ \sum_{n} {\mathfrak j}_{\,n}(\tau)\, e^{\,i\,\,n\xi }  \,.   
\end{equation}	
	For example, in the simplest case (\ref{LIE}) amplitude ${\mathfrak J}(\tau ,\xi)$
	satisfies the linear equation ($A=1$) \cite{Tal_EJM} 
		\begin{equation}
        \label{lin_eq_com}
\partial_\tau {\mathfrak J} =  - i  \partial_\xi^{\,2} {\mathfrak J}  
  - \frac{i}{2}\Bigl({\mathfrak J}   - \overline{\,{\mathfrak J}\,}\, \Bigr)   \,.
\end{equation}
Consequently,	
	\begin{equation}
	\label{sol_1}
        				{\mathfrak J}(\tau ,\xi) ~=~ \sum_{n} {\mathfrak j}_{\,n}\, 
								e^{\,i\,[\,n\xi  +  n \sqrt{n^2 - 1}\,\tau\,]} \,,  
\end{equation}	
	where the   amplitudes  ${\mathfrak j}_{\,n}$  satisfy the  condition 
		~$\overline{\,	{\mathfrak j}\,}_{\,-n}     ~=~    2 \left[n\sqrt{n^2 -1} - n^2 + \frac{1}{2} \right]  {\mathfrak j}_{\,n} \,.$
		
	Dynamics
	\[ {\mathfrak j}_{\,-n} ~\to~ {\mathfrak j}_{\,-n}(\tau) ~=~
		{\mathfrak j}_{\,-n}e^{\,i\, n \sqrt{n^2 - 1}\,\tau\,}\]
	is generated (for the conditional time  $t = t_0\tau$)  by Hamiltonian 
	\begin{equation}
  \label{energy_1}
 {H}_0(p_1,p_2,p_3\,;{\sf j}) ~=~ \frac{{\boldsymbol{p}}^{\,2}}{2m_0}   ~+~  
 {\mathcal E}_0 \sum_{n>1} |\,{\sf j}_{\,-n}|^2 n\sqrt{n^2 -1}	\,
\end{equation}  
	and Poisson brackets
	\begin{eqnarray}
  \{p_i\,,q_j\} & = & \delta_{ij}\,,\qquad i,j = x,y,z\,, \nonumber \\[2mm]
  \label{ja_jb_lin}
  \{ {\sf j}_{\,m}, \overline{\,\sf j\,}_{\,n}\} & = & (i/m_0 R_0 v_0)\, \delta_{mn}\,, \qquad m,n = -1,-2,\dots
  \end{eqnarray}
	In the  general  case  (\ref{New_H}), we have more complex but also linear equations for the amplitudes ${\mathfrak J}(\tau ,\xi)$.

	We assume  that the non-exited filament $\boldsymbol{r}_0(\tau ,\xi)$\,
	(see Eq. (\ref{proj_r}), where 	${\boldsymbol{j}}(\xi) \equiv {\boldsymbol{j}_0}(\xi)$), is described by the finite numbers of additional\footnote{to the variables  ${\boldsymbol j}_p(\tau ,\xi)$} dynamical variables, except for ''external'' variables ${\boldsymbol{p}}$ and ${\boldsymbol{q}}$. 
To simplify our subsequent studies, we consider the non-exited filament as a ring of radius $R$.
In case of the  Eq. (\ref{LIE_dim}), the following formula for the filament $\boldsymbol{r}_0(\tau ,\xi)$
is fulfilled:
\begin{equation}
        \label{our_sol}
 \boldsymbol{r}_0(\tau ,\xi) ~=~ \boldsymbol{q} + R
\bigl(\, \cos(\xi +\phi_0 +\beta\tau)\boldsymbol{e}_x\,, ~\sin(\xi +\phi_0+\beta\tau  )\boldsymbol{e}_y\,,  ~A \tau\boldsymbol{e}_z  \,\bigr)\,, 
\end{equation}
where the angle   $\phi_0 \in [0, 2\pi)$ and $\beta = B/R$.
Therefore, the variables
\[ \Delta R ~=~ \sqrt{R^2 - R_0^2}\,,\qquad \phi(\tau) = \phi_0 + \beta\tau \]
can be considered as the additional variables.
 Because the inequality $R \ge R_0$ is fulfilled, the constant  $R_0$ can  characterize the average size of a molecular cluster in a flow.

	To simplify our consideration,  we consider here the quantization of the variables 
	${\boldsymbol{p}}$,  ${\boldsymbol{q}}$ and ${\boldsymbol j}_p(\tau ,\xi)$ only. As regards of the quantization of the additional variables that define the filament
	$\boldsymbol{r}_0(\tau ,\xi)$, this procedure was fulfilled in the articles \cite{Tal_PoF23,Tal_PhRF23}  in simplest case (\ref{our_sol}).

	The constructed  set  of the classical dynamical variables defines the quantization scheme uniquely. The quantum states of a single vortex loop are the vectors of the Hilbert space
\begin{equation}
	\label{space_quant}
	\boldsymbol{H}_1  ~=~  \boldsymbol{H}_{pq}  \otimes   \boldsymbol{H}_j \,,
	\end{equation}
			where the symbol   $\boldsymbol{H}_{pq}$  denotes the Hilbert space  of a free structureless particle in $3D$ space ${\sf R}_3$.  As usually, we suppose that  $\boldsymbol{H}_{pq} = L^2({\sf R}_3)$.  				
			The	symbol $\boldsymbol{H}_j $ denotes the space for the  number of the harmonic oscillators  (infinite in general)  which correspond to the variables ${\mathfrak j}_{\,-n}$.
	In order to simplify the following formulas, 
		we assume  that  a finite number $K$  of modes ${\mathfrak j}_{-k}$ can be excited only for every vortex filament.								
					From the author's point of view, the proposed quantization scheme is a mirror image of Lord Kelvin's old idea that particles can be considered as some kind of vortex-like structures  \cite{Thom}. In our approach, we describe the closed vortex-like filaments  as  structured particles.

	As it is mentioned above, our quantization scheme make it possible 		
		to describe the interaction of vortex-like structures  in  terms of
 the   many-body theory.  The details can be found in the articles \cite{Tal_PoF23,Tal_PhRF23} for certain special case. 
Here  we will go the brief overview only.
\begin{itemize}
\item The space of a quantum states of a ''vortex tangle'' is constructed as the Fock space 
\[{\mathfrak H}  ~=~  \bigoplus_{M=0}^{\infty}\boldsymbol{H}_M \,,  \] 
where
\begin{equation}
\label{H_M}
 \boldsymbol{H}_M ~=~  \bigotimes_{i=0}^M \boldsymbol{H}_1 ~=~
 \underbrace{\,\bigl[\boldsymbol{H}_{pq}  \otimes   \boldsymbol{H}_j\bigr] \otimes  \dots \otimes \bigl[\boldsymbol{H}_{pq}  \otimes   \boldsymbol{H}_j\bigr]\,}_{M} =  \boldsymbol{H}_{pq}^M  \otimes   \boldsymbol{H}_j^M \,.
\end{equation}
Any vector  $|\Phi^M\rangle \in  \boldsymbol{H}_M$ has a form
 
\begin{eqnarray}
\label{vect_Phi}
|\Phi^M\rangle  &=&  \sum_{n_1,\dots,n_M} \int\,\cdots\int d\boldsymbol{p}_1\dots\boldsymbol{p}_N \,\times\nonumber\\[2mm]
~&\times&\,f^M_{{\mathfrak n}_1,\dots,{\mathfrak n}_M} (\boldsymbol{p}_1,\dots,\boldsymbol{p}_M)\, 
 |\boldsymbol{p}_1\rangle \dots|\boldsymbol{p}_M\rangle  | {\mathfrak n}_1\rangle\dots|{\mathfrak n}_M\rangle\,,
\end{eqnarray}

where the functions $f^M$ are symmetrical both indexes ${\mathfrak n}$ and arguments $\boldsymbol{p_\ell}$.  The vector  $|\boldsymbol{p}_\ell\rangle$ are  eigenvector of the operator $ \hat{\boldsymbol{p}}_\ell$ that corresponds to the filament with index $''{\ell}\,''$.
Here and further, we use the summation index $''{\ell}\,''$ to number the vortex filaments only.
As regards of the  ''numbers'' ${\mathfrak n}_\ell$, these symbols are reduced form for 
occupation numbers $n_k({\ell})$:
\[{\mathfrak n}_\ell ~=~  [\,n_{1}(\ell)\, \dots, n_{K}(\ell)\,]\,.\] 
The occupation number $n_{k}(\ell)$ describes the excitation of the variable ${\mathfrak j}_{-k}$ (see Eq. (\ref{sol_general})) for the filament with the index $''\ell\,''$.
\item   Hamiltonian
\begin{equation}
\label{ham_q_full}
\hat{\mathcal H} ~=~   \hat{\mathcal H}_0 +   \hat{U}\,, \nonumber
\end{equation}
where operator $\hat{\mathcal H}_0$ means the Hamiltonian for the system of the non - interacting vortex-like structures. 
\end{itemize}

 To write down these operators, let's first introduce notation for some operators in the Fock space ${\mathfrak H}$:
\begin{enumerate}
\item The creation and annihilation operators  ${\hat a}^+(\boldsymbol{p},{\mathfrak n})$,  ${\hat a}(\boldsymbol{p},{\mathfrak n})$  act in the Fock  space ${\mathfrak H}$.  
They are defined in a standard way:
\begin{equation}
 \Bigl({\hat a}(\boldsymbol{p},{\mathfrak n}) f^M_{{\mathfrak n}_1,\dots,{\mathfrak n}_M}  \Bigr)_{{\mathfrak n}_1,\dots,{\mathfrak n}_{M-1}}(\boldsymbol{p}_1,\dots,\boldsymbol{p}_{M-1}) ~=~
 \sqrt{M} f^M_{{\mathfrak n}_1,\dots,{\mathfrak n}_{M-1},{\mathfrak n}}
(\boldsymbol{p}_1,\dots,\boldsymbol{p}_{M-1},\boldsymbol{p})\,,\nonumber
\end{equation}
\begin{equation}
\Bigl({\hat a}^+(\boldsymbol{p},{\mathfrak n}) f^M_{{\mathfrak n}_1,\dots,{\mathfrak n}_M}  \Bigr)_{{\mathfrak n}_1,\dots,{\mathfrak n}_{M+1}}(\boldsymbol{p}_1,\dots,\boldsymbol{p}_{M+1}) ~=~{\hspace{57mm}}\nonumber
\end{equation}
\begin{equation}
 ~=~\Bigl({1}/{\sqrt{M+1}}\Bigr)\sum_{j=1}^{M+1}\delta_{{\mathfrak n},{\mathfrak n}_j}\delta(\boldsymbol{p} - \boldsymbol{p}_j)f^M_{{\mathfrak n}_1,\dots, \not{\mathfrak n}_j, \dots,  {\mathfrak n}_{M+1}}
(\boldsymbol{p}_1,\dots, \not\boldsymbol{p}_j \dots, \boldsymbol{p}_{M+1})\,.{\hspace{5mm}}\nonumber
\end{equation}
 As usual, the crossed-out symbol means its absence.
 \item Operators ${\hat a}^+_{m}(\ell)$,
${\hat a}_{m}(\ell)$  act in every Hilbert space $\boldsymbol{H}_j^M$. 
 These operators create or annihilate the oscillator energy level for the mode ${\mathfrak j}_{\,-m}$ in  the $\ell$-th   multiplier  $\bigl[\boldsymbol{H}_{pq}  \otimes   \boldsymbol{H}_j\bigr]$ of decomposition (\ref{H_M}). For example, the standard quantization rule for the value $|{\mathfrak j}_{\,-m}|^2$  looks like this:
\[|\,{\mathfrak j}_{\,-m}\,|^2 ~\to~ \frac{\hbar}{2}\left[\,{\hat a}^+_{m}(\ell){\hat a}_{m}(\ell) +  {\hat a}_{m}(\ell){\hat a}^+_{m}(\ell)\,\right]\,.\]
\end{enumerate}

Thus, summarizing of our consideration, we can write down a general formula for the operator  $\hat{\mathcal H}_0$:
\[\hat{\mathcal H}_0 ~=~   \bigoplus_{M=0}^{\infty}\hat{\mathcal H}_0^M\,,\]
where operators $\hat{\mathcal H}_0^M$ act in the spaces $\boldsymbol{H}_M$. 
These operators have following form:

\begin{eqnarray}
\label{ham_M_gen}
~&~&~\hat{\mathcal H}_0^M ~=~   \frac{\hbar^2}{2 m_0}\sum_{{\mathfrak n}}\int \boldsymbol{p}^2  {\hat a}^+(\boldsymbol{p},{\mathfrak n}){\hat a}(\boldsymbol{p},{\mathfrak n}) ~+\\[2mm]
~&+&\frac{\hbar}{2t_0} \sum_{\ell=1}^M\sum_{m=1}^K\Omega_{m \ell}(\varkappa_\ell\,; \alpha_1(\ell),\dots,\alpha_{\sf N}(\ell))  \,\left[{\hat a}^+_{m}(\ell){\hat a}_{m}(\ell) +  {\hat a}_{m}(\ell){\hat a}^+_{m}(\ell)\right]\,. \nonumber
\end{eqnarray}
The function $\Omega_{m\ell}(\varkappa_\ell\,; \alpha_1,\dots,\alpha_{\sf N})$ corresponds to the vortex-like structure with number $\ell$ in the considered vortex tangle.
Let us recall that the number ${\sf N}$ determines the level of our model (see Eq. (\ref{number_kappa})).
The explicit form of the functions $\Omega_{mn}(\varkappa_n\,; \alpha_1,\dots,\alpha_{\sf N})$ can be deduced from the detail study of linearized dynamical equations 
 (\ref{dyn_eq_g}).
For example, let the number ${\sf N}$  is equal to one  and corresponding number  $b_0 $ is also equal to one:   ${\sf N} = 1$, $b_0 =1$. This simplest case corresponds to
  a vortex tangle that  contains vortex loops described by  the equation  (\ref{LIE}).  
	The meaning of the constants $\alpha_1(\ell)$, $i=1,\dots,M$ is as follows:
	$\alpha_1(\ell) \propto |\boldsymbol{v}(\ell)|$, where the vector $\boldsymbol{v}(\ell)$ is the velosity of the vortex loop that numbered by index $'' \ell\,''$.
	The operator $\hat{\mathcal H}_0^M$ has the following  form:
\begin{eqnarray}
\label{ham_free1}
\hat{\mathcal H}_0^M &= &  \frac{1}{2 m_0 M}\sum_{{\mathfrak n}}\int \boldsymbol{p}^2  {\hat a}^+(\boldsymbol{p},{\mathfrak n}){\hat a}(\boldsymbol{p},{\mathfrak n}) ~+ \nonumber\\[2mm]
~&+& \frac{\hbar}{2t_0}   \sum_{\ell = 1}^M \sum_{k=1}^K \alpha_1(\ell)
k \sqrt{k^2 - 1}      \, \left[{\hat a}^+_{k}(\ell){\hat a}_{k}(\ell) +  {\hat a}_{k}(\ell){\hat a}^+_{k}(\ell)\right]\,. 
\end{eqnarray}
Let this  considered simplest vortex tangle is located in a closed  bounded domain $D \subset {\sf R}_3$.
Then, the eigenvalue promlem
\[ \hat{\mathcal H}_0^M|\Phi_{[\sigma]}^M\rangle ~=~    E_{[\sigma]}|\Phi_{[\sigma]}^M\rangle\,,
\qquad \quad  |\Phi_{[\sigma]}^M\rangle \in {\boldsymbol H}_M\,,   \]
has following solution. The eigenvectors $|\Phi_{[\sigma]}^M\rangle$ have form
(\ref{vect_Phi}), where 
\begin{eqnarray}
~&~&~f^M_{{\mathfrak n}_1,\dots,{\mathfrak n}_M} (\boldsymbol{p}_1,\dots,\boldsymbol{p}_M) ~=~
\frac{1}{(M!)^2}\times \nonumber\\[2mm]
~&\times &~\left(\sum_{\lfloor\boldsymbol{ p}_1,\dots,\boldsymbol{ p}_M\rceil}
\prod_{\ell=1}^M f_{w(\ell)} (\boldsymbol{ p}_\ell)\delta\bigl(\boldsymbol{ p}_\ell^2 - \lambda_{w(\ell)}^2\bigr)\right)
\left(\sum_{\lfloor{\mathfrak n}_1,\dots,{\mathfrak n}_M\rceil}\prod_{\ell=1}^M 
\delta_{{\mathfrak s}_\ell {\mathfrak n}_\ell }\right)\,,\nonumber
\end{eqnarray}

where the symbol  $\lfloor c_1,\dots,c_M\rceil$   means some rearrangement of the corresponding values  $c_1,\dots,c_M$   (summations are performed on all rearrangements).  Symbols ${\mathfrak s}_\ell$ mean the reduced form for certain  set of the occupation numbers $s_k({\ell})$:
\[{\mathfrak s}_\ell ~=~  [\,s_{1}(\ell)\, \dots, s_{K}(\ell)\,]\,.\] 
The function $f_{i}(\boldsymbol{p})\delta(\boldsymbol{p}^2 - \lambda_{i}^2)$ is the momentum representation of the
   eigenfunction ${\tilde f}_{i}(\boldsymbol{q})$  which corresponds to certain  eigenvalue
	$\lambda_i^2$   for the Laplace operator $\Delta$  in the domain $D$:
\[ \Delta {\tilde f}_{i}(\boldsymbol{q})   =  -\lambda_i^2 {\tilde f}_{i}(\boldsymbol{q})\,, \qquad  \boldsymbol{q} \in D\,, \quad {\tilde f}_{i}(\boldsymbol{q})\Big\vert_{\boldsymbol{q} \in \partial D} =~ 0\,.\]
Formula (\ref{ham_free1}) allows us to find eigenvalues:
\begin{eqnarray}
\label{E_simp}
E_{[\sigma]}\bigl(\alpha_1(1),\dots,\alpha_1(M)\bigr) &=& \frac{\hbar^2}{2m_0}\sum_{\ell=1}^M \lambda_{w(\ell)}^2  ~+~ \nonumber\\[2mm]
~&+& \frac{\hbar}{t_0}   \sum_{\ell = 1}^M \sum_{k=1}^{K} \alpha_1(\ell) \, k \sqrt{k^2 - 1}
\left( s_k(\ell) + \frac{1}{2}\right)\,.
\end{eqnarray}
To simplify the writing of these formulas, we have introduced a notation for the multi-index $[\sigma]$: 
\[[\sigma] ~=~  \bigl[\, w(1),\dots, w(M); s_1(1),\dots,s_{1}(M); \dots; s_K(1),\dots,s_{K}(M)\,\bigr]\,.\]
Let's analyze the structure of formula (\ref{E_simp}). The terms, which are proportional to the value of $\hbar$, give the main contribution to the energy. These terms represent the energy of oscillatory modes of the vortex rings. Terms that are proportional to the value of $\hbar^2$ can be interpreted as certain ''fine structure'' of oscillatory levels. This structure is completely determined by the geometry of the domain in which the considered vortex system is moving.

 The interaction Hamiltonian  $\hat{U}$ is written as follows:
\begin{equation}
\label{hat_U}
\hat{U} ~=~  \sum_{m,n } \varepsilon_{mn} \hat{U}_{m\leftrightarrow n}\,.
\end{equation}

The constants $\varepsilon_{m n}$ are  the coupling constants which define  the  intensity of the vortex  interaction.
Operators  $\hat{U}_{m\leftrightarrow n}$ define the  reconnection of the considered vortex filaments in the flow.  
They are constructed in terms of the creation and annihilation operators  ${\hat a}^+(\boldsymbol{p},{\mathfrak n})$  and  ${\hat a}(\boldsymbol{p},{\mathfrak n})$.  
Operator  $\hat{U}_{m\leftrightarrow n}$  describes the transformation of  $m$ vortex rings into $n$  rings and vice versa, $n \to m$.  For example, the theory with $U = \varepsilon\, \hat{U}_{1\to 2} $ was investigated in the paper  \cite{Tal_PoF23}, where the model  of the turbulent flow in it's initial stage was considered.
 We must emphasize that the 
variables ${\boldsymbol p}$ and  ${\boldsymbol q}$ describe the  virtual particles corresponding to the external degrees of freedom of the vortex-like loops. 
Therefore, the Hamiltonian $\hat{U}$ describes, generally speaking, a non-local interaction. 
These questions, as well as  corresponding examples of the operator $\hat{U}_{2\leftrightarrow 2}$,   were discussed in the article \cite{Tal_PhRF23}.
In this study, we will not dwell on this in detail.

	{\section{Analysis and outlooks}}

		~~~Despite the presence of a number of uncertain parameters, the proposed theory represents certain  new viewpoint to the description of turbulent flow. Some of these parameters appear to be random numbers. Apparently, this is unavoidable when describing a turbulence.
				Nevertheless, the approach developed in this study offers  a new look at the relationship between ''random'' and ''integrable'' dynamics in the description of a  fluids motion.
		According to the author, the proposed model can stimulate a vortex tangle even in the case of $U =0$,   especially for the large number of ${\sf N}$. Of course, the case where $U \not=0$ is more interesting and realistic.

	The proposed theory makes it possible to calculate the
partition function  ${\mathcal Z}$ for any concrete domain $D$. 
In the simplest case considered above,  we can write the following expression:
$$  {\mathcal Z}  = \prod_{\ell =1}^M \sum_{n}  \exp\left(- \frac{\langle E_{n}(\ell)\rangle }{{\sf k}_B{\sf T}}\right)\,, $$
where the numbers $E_{n}(\ell)$ are valid energy values for the vortex filament with number
$''\ell\,''$, symbol $\langle\dots\rangle$ means some averaging over the parameters $(\,\alpha_1(\ell),\dots, \alpha_{\sf N}(\ell)\,)$, 
 the value ${\sf T}$ is the  temperature and  the constant ${\sf k}_B$ is  Boltzmann constant. 	
		{  This possibility is an important feature of the proposed approach from a practical viewpoint. Indeed, the function ${\mathcal Z}$  allows us to calculate explicitly the thermodynamic characteristics of the simulated turbulent flow. }
			
			It is interesting to consider the limits  $K\to\infty$ and ${\sf N}\to\infty$.
			It is clear that the proposed model, within such limits, will require the application of certain quantum field theory methods. For example,  expression  (\ref{E_simp}) for the energy levels would require a renormalization procedure. 				
			The limit ${\sf N}\to\infty$  leads, in general, to the non-local equations for the single vortex-like structures.   As a cosequense, we have an additional  possibility to describe the non-local effects. 				In this case, the number  $\varkappa$ (see Eq.(\ref{number_kappa})) takes   any values in the segment  $[0,1]$.   	Note that non-local terms in LIE equation  appear when describing the self-stretching \cite{Majda}. 					
		Moreover there is another option to introduce non-local effects here: we can consider the additional non-local integrals  $I_{-n}$, $n = 1,2,3,\dots$ which can be introduced \cite{TakFad} for the 
		non-linear Schr\"odinger equation.  All these issues, as well as the physical interpretation of the proposed dynamical systems, require separate studies.
		
		\vspace{5mm}

	\section{Novelty, limitations and possible applications}
		
		Firstly, the proposed quantization scheme  leads to the richer circulation spectrum than the standard one. 
		This issue is studied  in the works of the author \cite{Tal_EJM,Tal_PoF22,Tal_PhRF23} in detail, so we do not discuss the relevant formulas here. From the author's point of view, this extention of a circulation spectrum  provides more opportunities for practical  modeling of a quantum turbulent flow.
		
		Secondly,  it is known that the standard approach to  the energy definition of a thin ($a\to 0$) vortex filament leads to problems due to the divergence of the integrals.	In this study, we applied a group-theoretical approach to the definition of energy.	
			
		Thirdly, as already noted above,	constructed quantum systems are ''inverse realization''  of Lord Kelvin's old idea that any particles can be considered as some kind of vortex-like structures  \cite{Thom}. In our approach, we describe the closed vortex-like filament  as some structured particle.  In a sense, the fluid in which the vortex  in question moves   plays the role of ''aether''.
	Note that such  ''aether'' is quite real here, and not hypothetical.

		The proposed quantization scheme for a system of vortex-like structures takes into account the movement of the surrounding fluid in a minimal way \cite{Tal_EJM}. 
	Apparently, this fact  imposes some limitations on the applicability of the constructed model. Undoubtedly, this issue requires further research.

\section{Concluding remarks}
 
{ The next stage in the development of the proposed model should be numerical simulation of some observable characteristics  of the flow  for   both various types of interaction (\ref{hat_U}) and various types of domain $D \subset {\sf R}_3$. }
 All these issues, as well as the  physical interpretation {and subsequent theoretical investigation} of the constructed dynamical systems, require separate studies.

		\end{document}